\documentclass[fleqn,usenatbib]{mnras}

\usepackage{newtxtext,newtxmath}

\usepackage[T1]{fontenc}
\usepackage[flushleft]{threeparttable}
\usepackage[font=normalsize]{caption}
\DeclareRobustCommand{\VAN}[3]{#2}
\let\VANthebibliography\thebibliography
\def\thebibliography{\DeclareRobustCommand{\VAN}[3]{##3}\VANthebibliography}

\usepackage{graphicx}	
\usepackage{amsmath}	
\usepackage{comment}

\title[Bars and Environment in Jellyfish Galaxies]{Stellar Bars in Jellyfish Galaxies: Statistical Insights into the Combined Role of Bars and Environment}

\author[Sánchez-García et al.]{%
Osbaldo Sánchez-García,$^{1,2}$\thanks{E-mail: osbaldo.schz@gmail.com}
Bernardo Cervantes Sodi,$^{2}$ \thanks{E-mail: b.cervantes@irya.unam.mx}
Jacopo Fritz,$^{2}$ 
Karín Menéndez-Delmestre,$^{1}$\newauthor 
Jacob P. Crossett, $^{3}$
Yasmin Cavalcante-Coelho $^{1}$  
\\
$^{1}$Universidade Federal do Rio de Janeiro, Observatorio do Valongo, Ladeira Pedro Ant\^{o}nio, 43, Sa\'{u}de CEP 20080-090 Rio de Janeiro, RJ, Brazil \\
$^{2}$Universidad Nacional Aut\'{o}noma de M\'{e}xico, Instituto de Radioastronom\'{i}a y Astrof\'{i}sica, Antigua Carretera a P\'{a}tzcuaro 8701, Ex-Hda, San Jos\'{e} de la Huerta,\\
58089 Morelia, Michoac\'{a}n, M\'{e}xico \\
$^{3}$ Departamento de F\'{i}sica, Universidad T\'{e}cnica Federico Santa Mar\'{i}a, Avenida España, 1680 Valpara\'{i}so, Chile
}

\date{Accepted XXX. Received YYY; in original form ZZZ}

\pubyear{\the\year{}}

\begin{document}
\label{firstpage}
\pagerange{\pageref{firstpage}--\pageref{lastpage}}
\maketitle

\begin{abstract}
Recent observational studies suggest that the interplay between internal and environmental mechanisms, in particular, the combined action of stellar bars and ram pressure stripping (RPS) may influence central star formation activity in jellyfish galaxies. However, current evidence relies on small samples, leaving open whether bars play a significant role during stripping. In this study, we analyse a sample about five times larger than those used in previous works, comprising 176 galaxies identified as RPS candidates based on optical morphological indicators such as asymmetries, debris tails, and displaced star-forming regions. To assess the impact of these processes, we examine radial $u-r$ colour profiles from SDSS imaging as tracers of the specific star formation rate (sSFR). We classify galaxies by bar presence and RPS signatures, and construct comparison samples through stepwise matching in stellar mass and environment to disentangle the individual and combined effects of bars and RPS on stellar population gradients. Our results show that central rejuvenation signals emerge in RPS candidate galaxies, becoming most evident when bars and RPS act together. Barred RPS galaxies are systematically bluer at all radii than their non-RPS counterparts, while unbarred systems display only mild or no central differences, suggesting that the observable outcome of RPS depends on the stripping stage. Furthermore, barred galaxies exhibit flatter central colour profiles than unbarred ones---a robust signature across all matched configurations. These findings highlight the key role of bars in amplifying environmental effects on the stellar populations of jellyfish galaxies, underscoring how internal structures can modulate the observable signatures of environmental processes in galaxies.

\end{abstract}




\section{Introduction} \label{sec: introduction}

Understanding the physical mechanisms that drive galaxy evolution remains a central question in astrophysics \citep{Somerville+2015}. Both internal processes, such as secular evolution driven by bars and radial gas inflows, and external environmental mechanisms, including galaxy interactions and ram pressure stripping (RPS), shape the morphology, dynamics, and stellar content of galaxies \citep{Gunn+1972, Kormendy+2004}.

Among the internal mechanisms of secular evolution, stellar bars are especially important. In nearby disc galaxies, optical visual classifications typically yield bar fractions of about 30–40\% when only strong or easily identifiable bars are considered \citep[e.g.][]{Masters+2011, Oh+2012, Ann+2015, Lee+2019}, while infrared surveys, which mitigate dust obscuration and are more sensitive to weak bars, report higher fractions up to about 60–70\% \citep[e.g.][]{Eskridge+2000, Menendez+2007}. The bar fraction also shows systematic correlations with global galaxy properties, increasing with stellar mass but being suppressed in blue, gas rich systems \citep{Masters+2012, CervantesSodi+2017, Erwin+2018}. On cosmological timescales, several studies report a decline in the bar fraction toward earlier epochs, with evidence up to $z \sim 1$ \citep{Sheth+2008, Melvin+2014, Huertas-Company+2025}. More recently, barred galaxies have also been identified at $z \sim 1$ to 3 \citep{Guo+2023, LeConte+2024}, and the bar fraction itself has now been measured out to $z \sim 4$, where it continues to show a decreasing trend with look back time \citep{Geron+2025}.

Bars strongly influence the internal evolution of disk galaxies. Their large-scale non-axisymmetric potentials promote angular momentum transfer between different components of the system \citep{Weinberg+1985, Athanassoula+1992, Sellwood+1993, Kormendy+2004}. By driving radial inflows, bars can funnel gas toward the central regions, sustaining enhanced star formation and producing bluer cores \citep[e.g.][]{Athanassoula+1992, Sheth+2005, Lin+2017, Chown+2019, Lin+2020}. Alternatively, they can redistribute material along their length, mixing stellar populations and inducing radial migration \citep[e.g.][]{Neumann+2020, Iles+2024, Bernaldez+2025}. In addition to these effects, bars can heat the stellar component and suppress star formation along their length through bar-induced quenching \citep[e.g.][]{Berentzen+1998, Athanassoula+2005, Berentzen+2007}, although recent observations have also reported ongoing or recent star formation along the bar \citep[e.g.][]{Fraser-McKelvie++2020, Neumann+2020, Bittner+2021}.

Observational studies further support the role of bars in regulating star formation. At the galaxy-integrated level, several large-sample analyses report that strongly barred systems can show lower present-day star formation rates than matched unbarred controls, in line with scenarios where earlier bar-driven inflows were followed by gas consumption \citep{Kim+2017, CervantesSodi+2017, Bitsakis+2019}. This interpretation is consistent with models in which inflows build central mass, promote bulge growth, and stabilise the disc, which reduces subsequent inflow efficiency \citep{Bournaud+2002, Athanassoula+2003}. Global trends are not purely bar driven, since galaxy interactions can enhance star-forming fractions and complicate relations between bars and star formation in population studies \citep{Ellison+2011, Wang+2012, Yoon+2020}. Taken together, these results indicate that the global impact associated with bars on current star formation varies with evolutionary stage and with the recent history of gas accretion and interactions.

On kiloparsec scales, spatially resolved analyses show that many barred galaxies exhibit renewed central star formation consistent with bar-driven inflows during the past 1 to 2 Gyr \citep{Ellison+2011, Wang+2012, Lin+2017, Lin+2020, Chown+2019}. The strength of this enhancement correlates with the concentration of central molecular gas, and the radius of the turnover correlates with bar length \citep{Chown+2019, Lin+2017, Lin+2020}. Other recent studies report centrally suppressed star formation and redder inner colours in barred systems, consistent with depletion and inside-out quenching driven by the bar \citep{Scaloni+2024, Renu+2025}. Central enhancements are also observed in galaxy pairs, which indicates that a bar can be sufficient to produce these signatures, although it is not necessary \citep{Chown+2019, Lin+2017}.

While the study of these processes is relatively straightforward in isolated systems, the situation becomes more complex in dense environments such as galaxy clusters. In these regions, galaxies can experience a variety of interactions, including gas stripping caused by RPS. This hydrodynamic mechanism acts when galaxies move through the high-density intracluster medium, producing shocks that can compress and partially or completely remove their interstellar gas \citep{Gunn+1972}.

As a result, RPS can lead to a depletion of the gaseous reservoirs of galaxies, ultimately quenching their star formation activity. Such signatures have been observed in clusters at low \citep{Merluzzi+2016, Poggianti+2017, Lopez-Gutierrez+2022}, intermediate \citep{Moretti+2022, Gibson+2025}, and high redshift \citep{McPartland+2016, Boselli+2019}. Depending on the galaxy mass and orbital configuration within the group or cluster, galaxies undergoing RPS can be observed in three main stages: initial stripping, peak stripping, and post-stripping \citep{Jaffe+2018, Poggianti+2025}.

The most extreme cases are the so-called jellyfish galaxies, where gas is being actively removed, producing prominent, one-sided gaseous tails that extend far from the stellar disk. In addition to gas removal, RPS can temporarily enhance star formation by compressing gas in the disk. This starburst phase has been reported in several studies \citep[e.g.][]{Roberts+2020, Lee+2022, Roberts+2022, Vulcani+2024, Lopez-Gutierrez+2025}. \citet{Poggianti+2016} found that galaxies likely affected by RPS lie above the star formation main sequence, indicating an excess of star formation. Similarly, \citet{Vulcani+2018} showed that ram-pressure affected galaxies exhibit enhanced star formation not only in their stripped tails but also across their disks.

This compression-driven activity boosts star formation for a limited time before the gas is fully removed, leading to quenching. In some cases, simulations suggest that the pressure gradients induced by RPS can also drive gas inflows toward the galaxy centre, potentially feeding transient nuclear activity or central star formation \citep[e.g.][]{Ramos-Martinez+2018, Akerman+2023, Kurinchi-Vendhan+2025}. These findings demonstrate that the influence of the hot intracluster medium can both intensify and redistribute star formation in disturbed disk galaxies, highlighting the need to consider this environmental effect when investigating the internal processes that regulate star formation.

Although the individual importance of stellar bars and RPS has been widely recognized, their combined effects on galaxy evolution have been little explored. Some recent studies have begun to investigate how the interaction between environmental pressure and the internal structure of galaxies may influence bar-driven gas flows. For instance, \citet{Bacchini+2023} analyzed four jellyfish galaxies undergoing RPS and reported radial molecular gas flows possibly related to bar-like instabilities. Similarly, \citet{SanchezGarcia+2023} used spatially resolved star formation histories of jellyfish galaxies and found that barred systems affected by ram pressure often show enhanced star formation and rejuvenated stellar populations in their central regions, particularly in galaxies observed near the peak of the stripping phase. These findings suggest that bars and ram pressure may act together to drive gas toward galaxy centres.

Despite these promising first efforts, a comprehensive assessment of how bars and ram pressure stripping interact to regulate central star formation across the broader cluster population is still lacking. Gaining a deeper understanding in this area is essential, as it connects two fundamental aspects of galaxy evolution: the internal processes that shape galaxies over long timescales and the external environmental factors that can rapidly alter their structure and star formation activity.

Building on the evidence from spatially resolved studies \citep[e.g.][]{SanchezGarcia+2023}, we extend the analysis to a statistically significant sample of cluster galaxies from Sloan Digital Sky Survey \citep[SDSS;][]{York+2000}. This allows us to investigate how the interplay between bars and ram pressure stripping influences central star formation across a broader and more diverse galaxy population. The aim of this paper is to provide a comprehensive assessment of these effects using a large, statistically controlled set of disc galaxies that combines bar classifications with signatures of ram-pressure stripping. By disentangling the roles of stellar mass, environment, and bars, we identify the conditions under which secular and environmental processes reinforce or counteract each other in shaping galaxy evolution.

This paper is organized as follows. Section \ref{sec: data} describes the data and sample selection. Section \ref{sec: bar_identification} details the identification of stellar bars, and Section \ref{sec: control_sample} presents the construction of control samples. Section \ref{sec: results} presents the main results and discussion. Finally, Section \ref{sec: conclusions} summarises the conclusions and their implications for galaxy evolution in cluster environments. Throughout, we adopt a standard flat cosmology with  $\Omega_{M}=0.3$, $\Omega_{\Lambda}=0.7$, and $H_{0}=70 \; \mathrm{km \; s^{-1} \; Mpc^{-1}}$.

\section{Data \& Sample selection} \label{sec: data}

In this study, we use two main galaxy samples. The first consists of RPS candidates compiled by \citet{Crossett+2025}, while the second is a volume-limited sample used as a comparison \citep{Lee+2012}. Galaxies in both samples were selected from the SDSS Data Release 7 \citep[DR7;][]{Abazajian+2009}, allowing us to obtain photometric, spectroscopic, and derived properties, as well as environmental information, through cross-matching with public value-added catalogs. For both samples, we adopt exactly the same redshift and absolute-magnitude limits defined in their original catalogues. These boundaries reflect the practical constraints of SDSS imaging, where the lower redshift cut helps minimise aperture effects, and the upper redshift and magnitude limits ensure sufficient spatial resolution for a reliable visual identification of morphological features.

\subsection{Ram pressure stripping sample}

The sample of RPS candidates is drawn from \citet{Crossett+2025} and comprises 313 galaxies with redshifts in the range $0.011 < z < 0.120$ and Petrosian absolute magnitudes in the $r$ band of $-21.88 < M_r < -16.07$. This compilation combines previously reported RPS systems from the literature with newly identified candidates selected through expert visual inspection of deep optical imaging.

\citet{Crossett+2025} first compiled a catalog of more than 900 galaxies previously reported in the literature as exhibiting morphological signatures consistent with ram pressure stripping, such as one-sided tails, debris, or compressed star-forming regions. From this compilation, 212 galaxies had morphological classifications available from Galaxy Zoo 2 \citep[GZ2;][]{Willett+2013} and suitable SDSS imaging, forming the baseline known RPS sample. These objects include galaxies with features identified across multiple wavelengths, including optical, UV, H$\alpha$, HI, and low-frequency radio continuum at 144 MHz \citep[see][for details]{Crossett+2025}.

Using the morphological trends identified in these 212 galaxies, such as a high fraction of “odd”, “irregular”, or “disturbed” classifications, \citet{Crossett+2025} then searched the full GZ2 database for galaxies with similar visual characteristics. Through expert visual inspection of $grz$ images from the DESI Legacy Imaging Surveys \citep{Dey+2019}, they identified 101 additional RPS candidates exhibiting optical features analogous to those seen in previously confirmed systems. Typical features include trailing debris, gas compression, and asymmetric star-forming regions, similar to those reported in previous studies \citep[e.g.][]{Poggianti+2016, Roberts+2020, Kolcu+2022, Piraino-Cerda+2024}.

It is important to note that, although most galaxies in this sample were compiled from the literature, their classification as RPS systems is based solely on visual inspection of morphologies consistent with RPS. For this reason, they should be considered candidates for experiencing RPS. While further multi-wavelength observations and kinematic analyses are required to confirm the presence of ongoing gas stripping, visual identification has been shown to be a reliable proxy for selecting RPS galaxies, with nearly 90\% of imaging-selected candidates confirmed by integral-field spectroscopy in the GASP survey \citep{Poggianti+2025}. The sample therefore includes both previously confirmed cases of gas stripping and new candidates identified through visual inspection; however, we do not differentiate between these subsets in our statistical analysis. Throughout this study we adopt the term RPS galaxies for clarity and consistency.

\subsection{Comparison Sample}

To assess the impact of stellar bars in galaxies affected by RPS, we use a comparison sample drawn from the volume-limited catalog of 33,391 galaxies presented by \citet{Lee+2012}, selected from SDSS DR7. This sample spans a redshift range of $0.020 \leq z \leq 0.055$ and Petrosian absolute magnitudes in the $r$-band of $-23.64 < M_r < -6.93$. It includes visual morphological classifications, as well as the identification of the presence or absence of stellar bars (see Section~\ref{sec: bar_identification}).

Because the comparison sample is volume-limited and spans a wide range of typical environments, it is expected to be dominated by galaxies not undergoing significant environmental processes. The occurrence of RPS in the local Universe is relatively rare, with estimated fractions of less than 5\% in general galaxy samples \citep[e.g.,][]{Poggianti+2016, Roberts+2021, Kolcu+2022}, and is mostly associated with dense cluster environments \citep[e.g.,][]{Gunn+1972, Roberts+2021}. Therefore, while we cannot entirely rule out the presence of RPS-affected galaxies in this sample, it serves as a reasonable proxy for a non-RPS population. We hereafter refer to this comparison sample as NRPS (non–ram pressure stripped) galaxies.

\subsection{Galaxy Properties}

We obtained the main photometric, spectroscopic, morphological, and environmental properties of our galaxy samples through cross-matching with a set of public value-added catalogs based on SDSS DR7.

\subsubsection{Photometric and Structural Properties}

Photometric and structural properties of the galaxies were obtained from two main sources: the Korea Institute for Advanced Study Value-Added Galaxy Catalog \citep[KIAS-VAGC;][]{Choi2010} and the NASA-Sloan Atlas \citep[NSA;][]{Blanton+2011}. Redshifts, absolute $r$-band magnitudes ($M_r$), axial ratios ($b/a$), and morphological classifications were taken from the KIAS-VAGC. This catalog is a refined version of the Large Scale Structure (LSS) sample of the New York University Value-Added Galaxy Catalog \citep[NYU-VAGC;][]{Blanton+2005}, based on SDSS DR7 data, and incorporates additional photometric corrections, morphological classifications, and improved completeness for low-redshift galaxies. The KIAS-VAGC also provides a robust classification into early- and late-type galaxies based on color gradients and light concentration, supplemented by visual inspection in ambiguous cases. We use the multi-band cutout images retrieved from the NSA, which offers reprocessed imaging with improved background subtraction, photometric calibration, and spatial coverage tailored to extended sources in the nearby universe.

After cross-matching the RPS and NRPS samples with the KIAS-VAGC and NSA catalogues, a total of 278 RPS and 31,498 NRPS galaxies were retained.

\subsubsection{Stellar Masses and Emission Line Properties}

Stellar masses and emission line measurements were obtained from the MPA-JHU catalog based on SDSS DR7 (www.mpa-garching.mpg.de/SDSS/DR7/). Stellar mass estimates are derived using the Bayesian methodology and model grids of \cite{Kauffmann++2003}, fitting the broadband SDSS ugriz photometry rather than spectral indices. The photometry is corrected for nebular emission using the spectra, and a \citet{Kroupa+2001} initial mass function is assumed. Emission lines are measured following the methodology of \cite{Brinchmann+2004}, where the stellar continuum is first subtracted using \cite{Bruzual+2003} population synthesis models, and the residual spectra are then fitted with Gaussian profiles. After cross-matching, a total of 277 RPS and 29,485 NRPS galaxies were retained.

Furthermore, to focus on bar- or RPS-driven effects on star formation activity, we excluded galaxies hosting an AGN, as their presence can contaminate central star formation tracers. AGN identification was carried out using the standard BPT diagnostic diagram \citep{Baldwin+1981}, a widely used method to classify galaxies according to their dominant ionization source, based on the emission-line ratios [O\,\textsc{iii}]/H$\beta$ versus [N\,\textsc{ii}]/H$\alpha$, adopting the demarcation criteria proposed by \citet{Kewley+2001} and \citet{Kauffmann+2003}. Galaxies lying above the \citet{Kewley+2001} line (pure AGN or LINERs) were excluded, while those classified as star-forming or located in the composite region between the \citet{Kauffmann+2003} and \citet{Kewley+2001} curves were retained, as they are still expected to host substantial star formation. After applying these criteria, the final sample comprises 245 RPS and 15\,059 NRPS galaxies used in the subsequent analysis.

\subsubsection{Environmental Properties} \label{sec: data_environment}

In this study, we use the phase–space position as a proxy for the environmental conditions of galaxies. We characterize the environment using the line-of-sight velocity offset from the cluster mean, normalized by the cluster velocity dispersion ($\Delta V_{\mathrm{cl}}/\sigma_{\mathrm{cl}}$), as a function of the projected clustercentric distance normalized by the virial radius ($r_{\mathrm{cl}}/r_{\mathrm{vir}}$).

These phase–space diagrams provide a convenient way to trace the orbital histories of galaxies within clusters and to infer their likelihood of undergoing environmental processes. In particular, the phase–space locus correlates with the probability of experiencing ram pressure stripping, with galaxies located closer to the cluster centre and moving at higher velocities being more likely to be affected \citep[e.g.][]{Jaffe+2015, Rhee+2017}.

Environmental information, including group membership and phase–space positions, was taken from the galaxy group catalogue compiled by \citet{Tempel+2014}. This catalogue was constructed using a modified friends-of-friends algorithm with a variable linking length, applied to the spectroscopic galaxy sample from SDSS Data Release 10 \citep{Ahn+2014}, and provides both flux-limited and volume-limited group and cluster identifications.

\begin{figure*}
\includegraphics[width=0.8\textwidth]{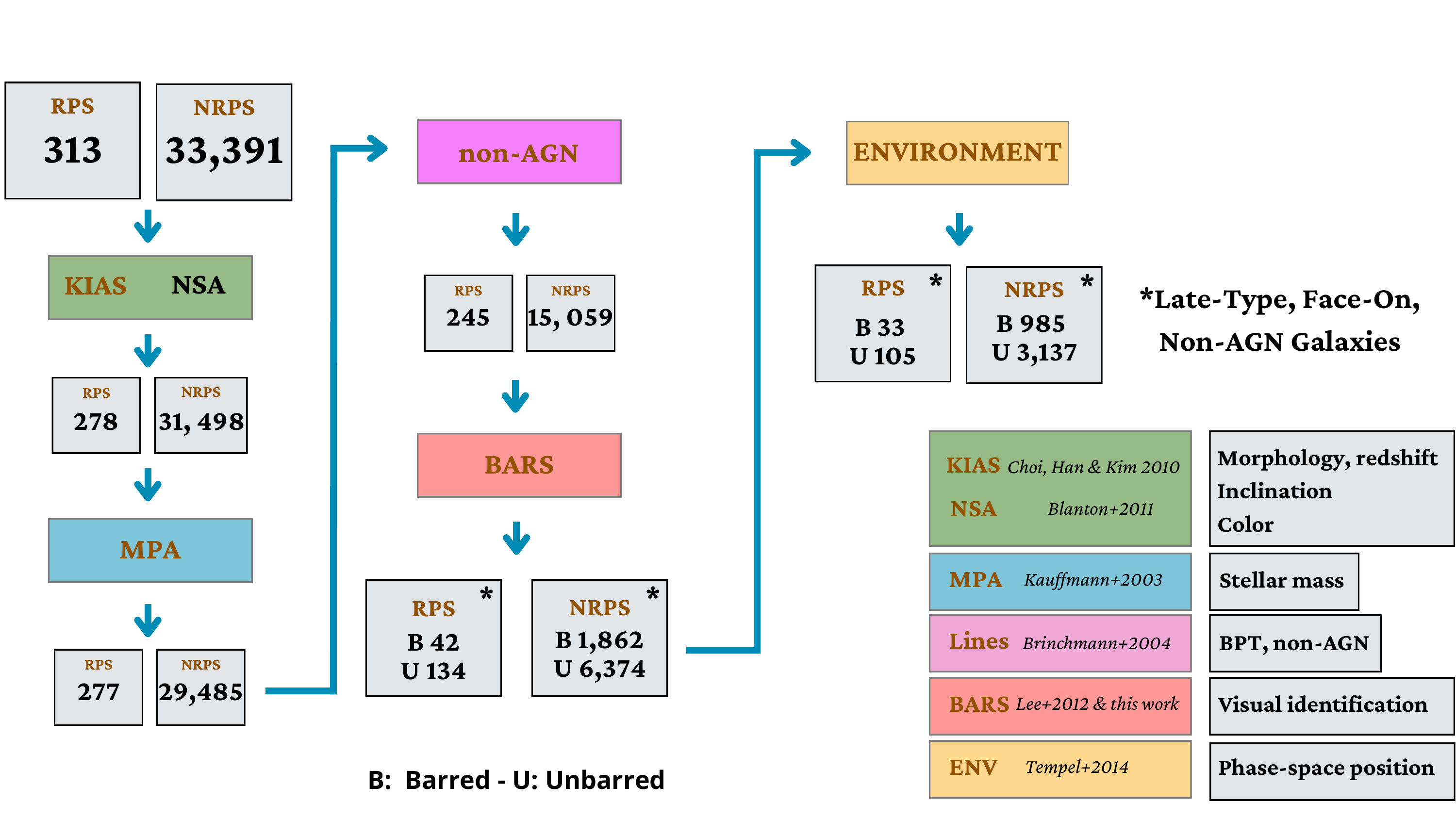}
    \caption{Workflow followed to construct the final galaxy samples. The process consists of successive cross-matches with SDSS public value-added catalogs, indicating the number of galaxies retained at each step. The final stage includes the bar classification (Section \ref{sec: bar_identification}), leading to the working samples of late-type, face-on galaxies without AGN. Acronyms (KIAS-VAGC, NSA, MPA-JHU) are defined in Section \ref{sec: data}, and further details on the bar classification procedure are provided in Section \ref{sec: bar_identification}.}
    \label{fig:workflow}
\end{figure*}

After performing the cross-match with the environmental catalogue, a total of 201 RPS and 8,096 NRPS galaxies with available phase–space position information were retained. All RPS galaxies are located within five virial radii ($R < 5R_\mathrm{vir}$) of their respective cluster centres, while only a small fraction of NRPS galaxies (13 out of 8,096, $\sim$0.2 per cent) extend slightly beyond this limit, reaching up to $\sim$6.3 $R_\mathrm{vir}$. All galaxies have line-of-sight velocity offsets below three times the cluster velocity dispersion ($|v|/\sigma < 3$). Figure~\ref{fig:workflow} summarises the entire sample construction process, including the bar classification described in Section~\ref{sec: bar_identification}, leading to the final subsets used in our analysis.

\begin{figure}
	\includegraphics[width=\columnwidth]{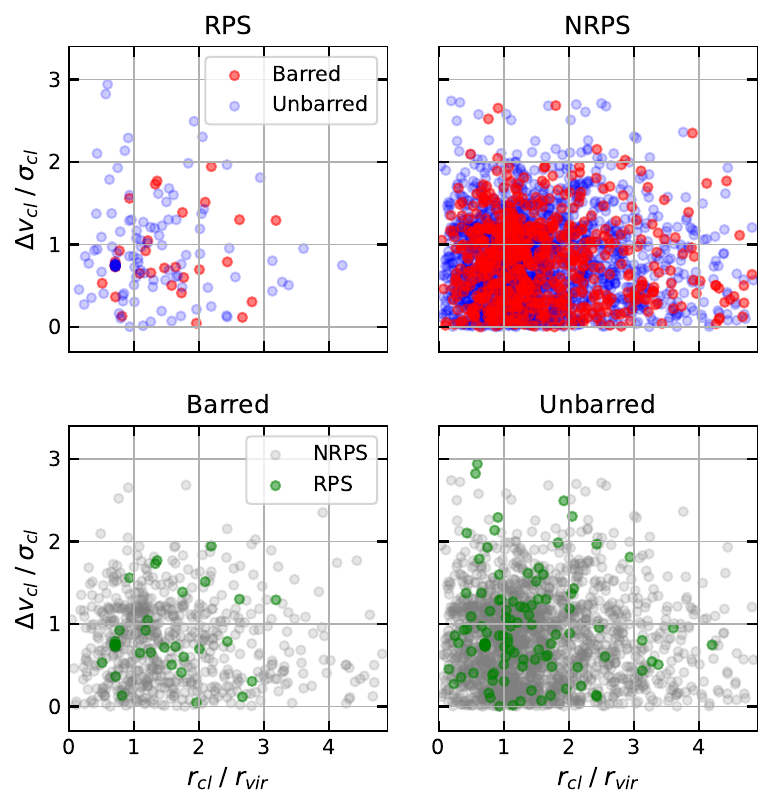}
    \caption{Projected phase–space diagram illustrating the distribution of galaxies  in $\Delta V_{cl}/\sigma_{cl}$ (line-of-sight velocity offset from the cluster mean normalized by the cluster velocity dispersion) versus $r_{cl}/r_{vir}$ (projected clustercentric distance normalized by the virial radius).  Top-left panel: galaxies affected by RPS, including both barred and unbarred types.  Top-right panel: galaxies from the comparison (NRPS) sample, categorized into barred and unbarred.  Bottom-left panel: barred galaxies split into RPS and NRPS subsamples.  Bottom-right panel: unbarred galaxies split in the same way.}

    \label{fig:pv_all}
\end{figure}

\section{Bar identification} \label{sec: bar_identification}

Stellar bars were identified through visual inspection of color-composite images, where color information facilitates the recognition of elongated central structures. We restricted our analysis to late-type galaxies without AGN activity, and further required them to be face-on ($b/a > 0.5$). This criterion, commonly adopted in previous studies \citep[e.g.][]{Li+2011, Masters+2011, Lee+2012}, minimizes projection effects and internal dust extinction that could obscure bar features in more inclined galaxies.

\begin{figure}
	\includegraphics[width=\columnwidth]{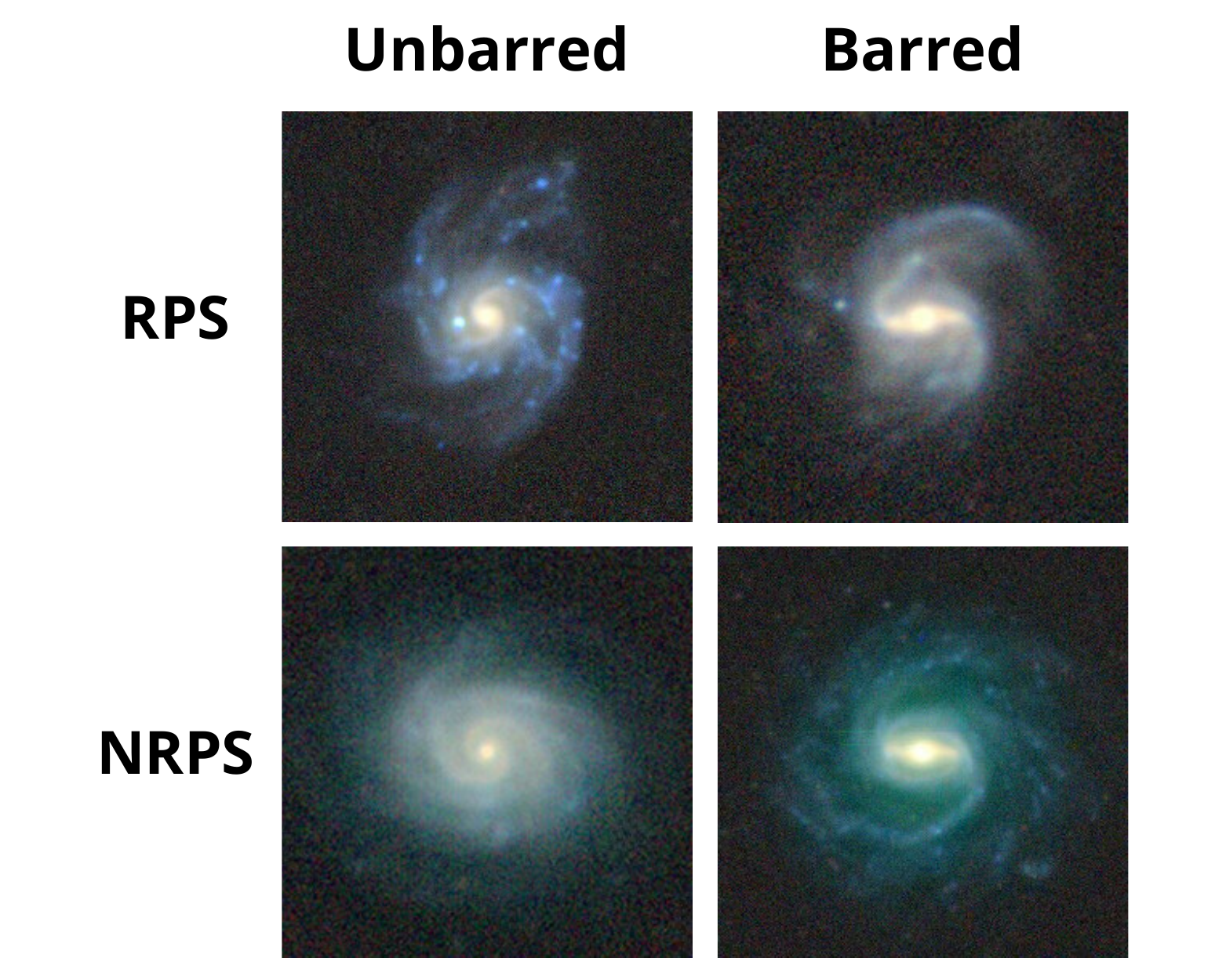}
        \caption{Representative examples of visually classified galaxies from our samples. The top row shows RPS galaxies and the bottom row NRPS galaxies, each split into unbarred (left) and barred (right) categories. Images are $g+r+z$ colour composites from the DESI Legacy Imaging Surveys (Data Release 9; \citealt{Dey+2019}).}
    \label{fig:bar_examples}
\end{figure}

For the NRPS sample, bar classifications were taken directly from \citet{Lee+2012}, who visually inspected SDSS $g+r+i$ composite images to categorize galaxies as barred or unbarred.  Although \citet{Lee+2012} further distinguished between strong and weak bars, we did not separate these categories, focusing instead on the simple presence of a bar.

In parallel with the \citet{Lee+2012} classification, for the RPS sample, we carried out our own visual classification using $g+r+z$ composite images from the DESI Legacy Imaging Surveys \citep{Dey+2019}.  Bars were identified when an elongated central structure, distinct from the bulge  and extending into the disk, was clearly recognizable. Three co-authors independently inspected the images, and ambiguous cases were resolved through discussion until consensus was reached, ensuring the robustness of the classification. Representative examples of barred and unbarred galaxies from both samples are shown  in Figure~\ref{fig:bar_examples}, illustrating the consistency of the visual classification across the two datasets.

After applying the selection criteria and bar classifications, we obtained the final samples of RPS and NRPS galaxies. All have complete structural and spectroscopic data, and a subset with available environmental information was used for the phase-space analysis. In total, we identified 42 barred and 134 unbarred galaxies in the RPS sample, and 1,862 barred and 6,374 unbarred in the NRPS sample. Of these, 33 barred and 105 unbarred RPS galaxies, and 985 barred and 3,137 unbarred NRPS galaxies include environmental information. Their distribution in phase space is shown in Figure \ref{fig:pv_all}.

\section{Control sample description} \label{sec: control_sample}

To identify potential signatures of the combined influence of stellar bars and RPS, we construct control samples that are statistically matched in key physical properties, specifically selected to minimize biases arising from differences in stellar mass and environment, across four galaxy subsets defined by the presence or absence of bars and RPS: (1) galaxies affected by RPS, further divided into barred and unbarred; (2) galaxies unaffected by RPS, also split into barred and unbarred; (3) barred galaxies, categorized by whether or not they are affected by RPS; and (4) unbarred galaxies, likewise classified based on RPS influence. This classification allows us to systematically assess the individual and combined effects of bars and RPS on the stellar populations of disk galaxies, ensuring that observed differences are not driven by underlying disparities in galaxy characteristics.

To implement this matching strategy, we use Propensity Score Matching (PSM) with the {\sc MatchIt} package \citep{Ho+2011} in R \citep{R+2015}, adopting the {\sc full} matching method. This approach retains all galaxies in the analysis and assigns a statistical weight to each one based on its similarity to galaxies in the opposite group. As a result, the matched samples become comparable in terms of key covariates, enabling unbiased comparisons between subsets. In this sense, because full matching does not remove galaxies, the matched subsets retain the same numerical sizes as the corresponding final samples (see Section \ref{sec: bar_identification}): after stellar–mass matching the samples contain 176 RPS and 8,236 NRPS galaxies, while the mass– and environment–matched subsets include 138 RPS and 4,122 NRPS galaxies. This choice is particularly important given the relatively small size of the RPS sample, where discarding galaxies would severely limit the statistical power of our analysis.

{\sc Full} matching operates by constructing matched sets (or subclasses) that contain at least one treated unit and one or more control units, or vice versa. It optimizes the assignment of units to these sets by minimizing the average distance in propensity score space between treated and control galaxies within each subclass \citep{Rosenbaum2002,Hansen2004}. This method is particularly well suited for situations involving multiple covariates, as it allows flexible subclass sizes, avoids excluding galaxies with rare properties, and achieves good covariate balance. The resulting weights are then incorporated into subsequent statistical analyses to properly account for the matched design.

To evaluate the balance achieved by the matching procedure, we perform Kolmogorov–Smirnov (KS) tests comparing the distributions of each covariate between treated and control galaxies in the matched sample. Values of the \textit{p}-statistic close to 1 indicate that the two samples are statistically indistinguishable. We implemented this approach to construct one control sample matched only in stellar mass and another matched simultaneously in stellar mass and environment, as described in the following subsections.

\subsection{Stellar mass-matched control sample}\label{control by mass}

As a first step, we perform full matching using stellar mass as the only covariate to ensure that comparisons between galaxy subsets are not biased by differences in mass distribution.

\begin{figure}
	\includegraphics[width=\columnwidth]{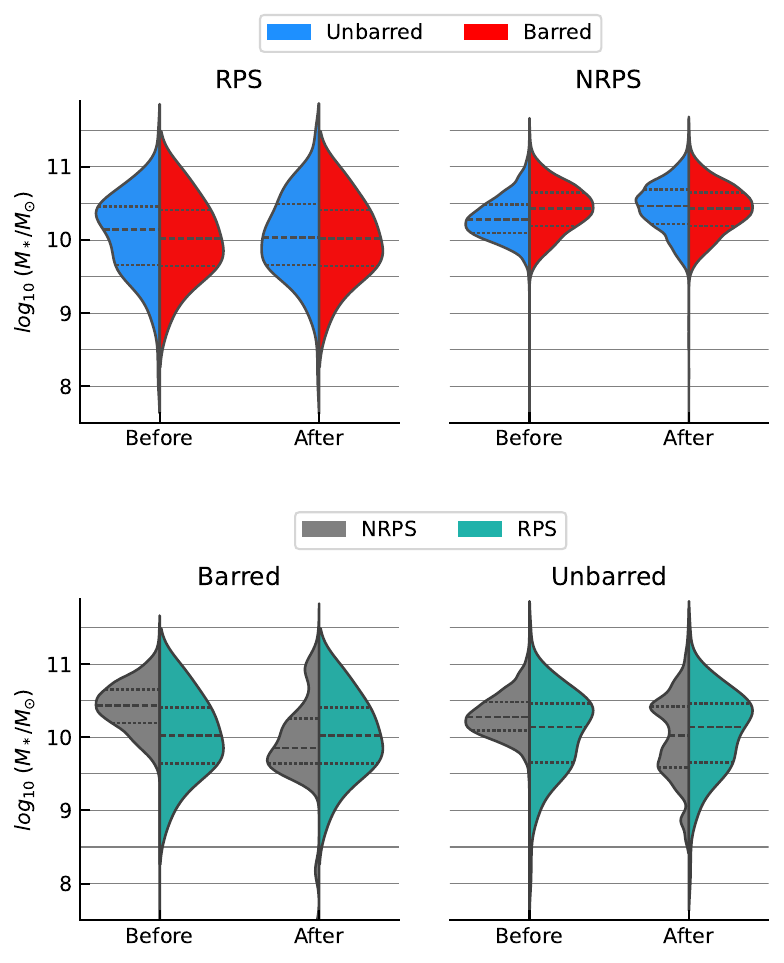}
    \caption{Violin plots illustrating the distribution of stellar masses in galaxies before and after the matching process. The {\it upper panels} distinguish between RPS and NRPS galaxies, further categorized into barred (red) and unbarred (blue) types. The {\it bottom panels} differentiate galaxies based on the presence of bars, further divided into barred (red) and unbarred (blue) galaxies.}
    \label{fig:violin_plot_mass}
\end{figure}

Figure~\ref{fig:violin_plot_mass} shows the distributions of stellar mass before and after matching, grouped according to RPS status and bar classification. The associated KS test \textit{p}-values are summarized in Table~\ref{tab:match_mass}. Before matching, significant discrepancies were present in all subsets, with \textit{p}-values equal to or near zero. After matching, all distributions reached \textit{p}-values of 1.0, indicating that they are statistically indistinguishable.

The resulting matched samples span a stellar mass range of approximately \( \log_{10}(M_\ast/M_\odot) \sim 7.3 \) to 11.6, with median values across subsets consistently around 10.2--10.4. These values confirm that the galaxies included after matching are distributed over a representative and consistent mass range, suitable for controlled comparisons across the defined subsets.

\begin{table}
\begin{threeparttable}
\begin{center}
\caption{\textit{P}-values from the Kolmogorov-Smirnov test for subset samples before and after the stellar mass matching process.}
\begin{tabular}{cc|cc}\hline \hline
    \multicolumn{4}{c}{Matched by $M_\ast$}\\ \hline
    \multicolumn{1}{c}{Subset} & \begin{tabular}[c]{@{}c@{}} Number of\\ galaxies\end{tabular} & \begin{tabular}[c]{@{}c@{}}KS before\\ matching\end{tabular} & \begin{tabular}[c]{@{}c@{}}KS after\\ matching\end{tabular} \\ \hline
     RPS      & 176   & 0.515 & 1.000\\
     NRPS     & 8,236 & 0.000 & 1.000\\
    Barred    & 1,904 & 0.000 & 1.000\\
    Unbarred  & 6,508 & 0.000 & 1.000\\
    \hline
    \hline
\end{tabular}
\label{tab:match_mass}
\begin{tablenotes}
\item \noindent \textbf{Notes:} KS test \textit{p}-values indicate whether the covariate distributions are statistically distinguishable before and after matching.
\end{tablenotes}
\end{center}
\end{threeparttable}
\end{table}

\subsection{Environment-matched control sample} \label{control by environment}

The next step involves constructing a control sample that takes both stellar mass and the environment into account. As discussed in Section~\ref{sec: data_environment}, to estimate the potential impact of RPS we consider galaxy positions in projected phase space, which serve as a proxy for their orbital stage within the cluster and, indirectly, for the intensity of the interaction with the intracluster medium.

\begin{figure}
	\includegraphics[width=\columnwidth]{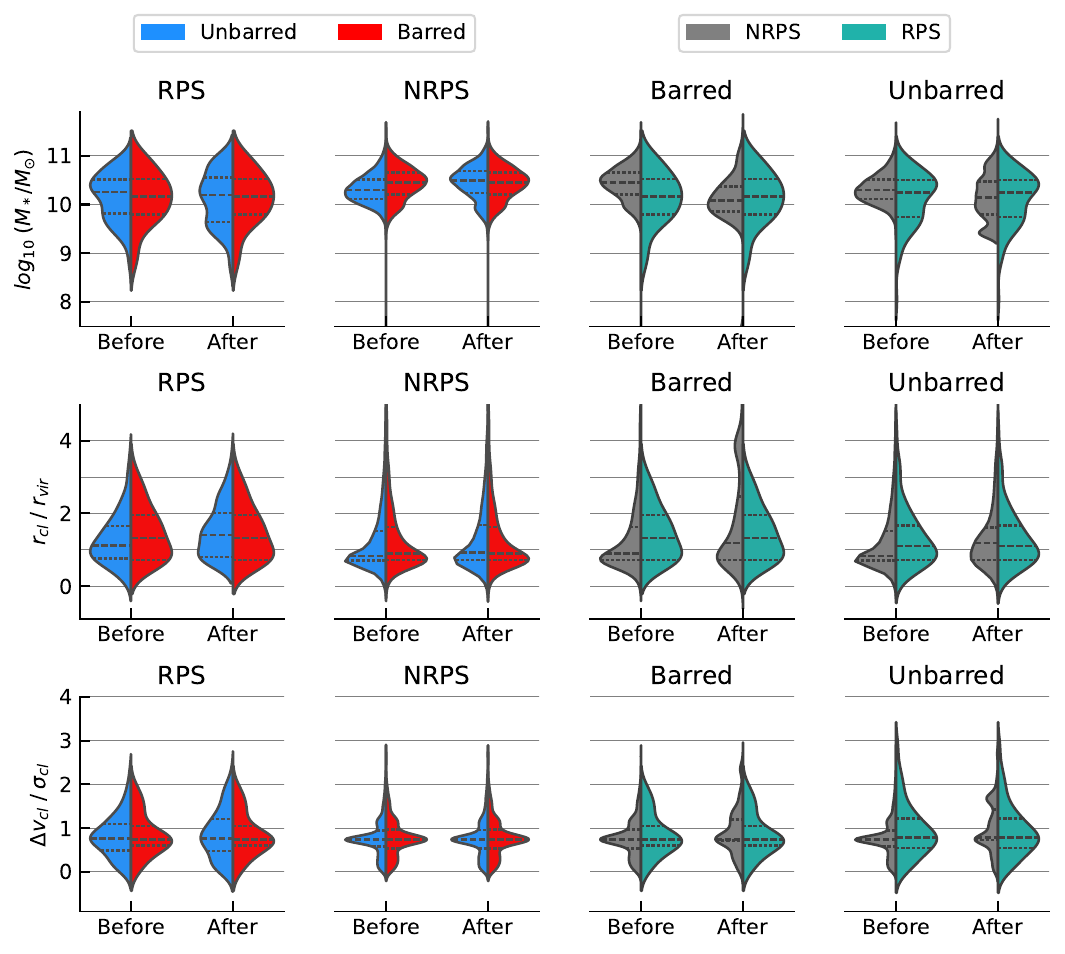}
    \caption{Violin plots illustrating the distribution of stellar masses ($M_\ast$), projected cluster distance ($r_{cl}/r_{vir}$), and line-of-sight velocity ($\Delta V_{cl} / \sigma_{cl}$) of galaxies before and after the matching process. The panels distinguish between galaxies affected by RPS and NRPS, further categorizing them into barred (red) and unbarred (blue) types. Additionally, galaxies are differentiated based on the presence of bars, with further subdivisions into those affected (green) and unaffected (black) by RPS.}
    \label{fig:violin_plot_env}
\end{figure}

Figure \ref{fig:violin_plot_env} illustrates the distributions of these parameters for the four subsets studied in this work. The upper panels display the results before and after matching for stellar mass, the middle panels report the distribution with respect to normalized clustercentric distance, and the lower panels illustrate the distribution of projected velocities normalized by cluster velocity dispersion. Table \ref{tab:match_environment} reports \textit{p}-values obtained from the KS test performed before and after the matching process.

\begin{table*}
\begin{threeparttable}
\begin{center}
\caption{\textit{P}-values  from the Kolmogorov--Smirnov (KS) test for each subset before and after matching by stellar mass and environmental parameters ($M_\ast$, $r_{\mathrm{cl}}/r_{\mathrm{vir}}$, and $\Delta v_{\mathrm{cl}}/\sigma_{\mathrm{cl}}$). }
\begin{tabular}{lccccccc}\hline
    \multicolumn{8}{c}{Matched by $M_\ast$, $\Delta v_{cl}/ \sigma_{cl}$, $r_{cl}/r_{vir}$ }\\ \hline \hline 
    & & \multicolumn{2}{c}{ $M_\ast$} & \multicolumn{2}{c}{ $r_{cl}/r_{vir}$}&\multicolumn{2}{c}{$\Delta v_{cl}/ \sigma_{cl}$}\\
     \hline
    \multicolumn{1}{c}{Subset} & \begin{tabular}[c]{@{}c@{}} Number of\\ galaxies\end{tabular} & \begin{tabular}[c]{@{}c@{}}KS before\\ matching\end{tabular} & \begin{tabular}[c]{@{}c@{}}KS after\\ matching\end{tabular}& \begin{tabular}[c]{@{}c@{}}KS before\\ matching\end{tabular} & \begin{tabular}[c]{@{}c@{}}KS after\\ matching\end{tabular}& \begin{tabular}[c]{@{}c@{}}KS before\\ matching\end{tabular} & \begin{tabular}[c]{@{}c@{}}KS after\\ matching\end{tabular} \\ \hline
RPS       & 138 & 0.913 & 0.997 & 0.497 & 1.000 & 0.239 & 0.966\\
NRPS      & 4,122 & 0.000 & 1.000 & 0.018 & 1.000 & 0.462 & 1.000\\
Barred    & 1,018 & 0.003 & 0.757 & 0.075 & 0.884 & 0.073 & 0.992\\
Unbarred  & 3,242 & 0.000 & 0.999 & 0.000 & 0.981 & 0.043 & 0.998\\
\hline
\end{tabular}
\label{tab:match_environment}
\begin{tablenotes}
\item \noindent \textbf{Notes:} KS test \textit{p}-values indicate whether the covariate distributions are statistically distinguishable before and after matching.
\end{tablenotes}
\end{center}
\end{threeparttable}
\end{table*}

The resulting matched samples span a stellar mass range of approximately \( \log_{10}(M_\ast/M_\odot) \sim 7.3 \) to 11.6, with median values around 10.2--10.4 across all subsets. In terms of projected distance, values range from \( r_{\mathrm{cl}}/r_{\mathrm{vir}} \sim 0.04 \) to 6.3, with medians near 0.86--1.14. For velocity, \( \Delta V_{\mathrm{cl}}/\sigma_{\mathrm{cl}} \) ranges from 0.0005 to 2.94, with median values consistently around 0.74 in all subsets, except for the RPS group, which shows a broader spread and no defined median due to its skewed distribution.

Although exact alignment across all parameters is more challenging when considering multiple covariates, the visual and statistical comparisons indicate that the matching procedure substantially improves balance. This justifies the use of these samples for controlled comparisons involving environmental conditions.

With these matched samples, we ensure that the distributions of stellar mass and key environmental parameters are statistically comparable across all galaxy subsets. This provides a robust basis to disentangle the relative and combined effects of stellar bars and ram-pressure stripping.

\section{Results and Discussion} \label{sec: results}

\subsection{Color radial profiles} \label{sec:color_profiles}

To determine whether star formation in disk galaxies is enhanced or quenched by the combined action of gas removal, compression, and redistribution, driven by both ram pressure from the hot intracluster medium and secular processes such as bar-driven inflows, we analyse their impact on the underlying stellar populations. Although these mechanisms act primarily on the gas, their cumulative effects can leave observable imprints on the spatial distribution of stellar populations. To trace such imprints, we rely on the $u-r$ colour as a proxy for the specific star formation rate (sSFR), defined as the star formation rate per unit stellar mass, which in turn reflects recent star formation activity across the disk. The $u$ band is sensitive to young stellar populations, while the $r$ band broadly traces the underlying stellar mass; therefore, their combination provides an effective photometric tracer of sSFR. This approach is motivated by the lack of spatially resolved spectroscopic data for large RPS samples, where broad-band photometry offers the only practical alternative.

We examine the radial variation of the $u-r$ colour across the four galaxy subsets defined by the presence or absence of bars and RPS (see Section~\ref{sec: control_sample}). Magnitudes in the $u$ and $r$ bands are measured in concentric annuli at different galactocentric distances, normalized to the Petrosian radius of each galaxy. Radial bins of 0.2 Petrosian radii are adopted to ensure consistent spatial resolution. The median $u-r$ value in each bin is used to build the radial profiles, with uncertainties estimated from the dispersion obtained via bootstrap resampling.

We begin by examining unconstrained profiles, and then progressively refine the analysis by controlling first for stellar mass and then for both stellar mass and environmental conditions. Rather than applying all constraints simultaneously from the start, this stepwise approach is designed to reveal how the $u-r$ color profiles change as we sequentially account for potential sources of bias. By comparing results at each stage, we can assess the extent to which stellar mass and environment influence the observed differences, and better isolate the individual and combined effects of bars and RPS. The results are presented in Figures~\ref{fig:rad_prof_all}, \ref{fig:rad_prof_mass}, and \ref{fig:rad_prof_mass_env}. In the NRPS panels, the bootstrap 1$\sigma$ dispersions are included in all radial bins. However, because the NRPS subsets contain several thousand galaxies, the resulting dispersions are comparable to or smaller than the marker size. The profiles are plotted as a function of the galactocentric radius, normalized by the Petrosian radius enclosing 50\% of the total flux, $R_{\mathrm{P,50}}$,  on a logarithmic scale to emphasize the central region, log($R/R_{\mathrm{P,50}}$) < 0.

\subsubsection{Unconstrained Color Radial Profiles} \label{subsec: unconstrained_crp}

\begin{figure}
	\includegraphics[width=\columnwidth]{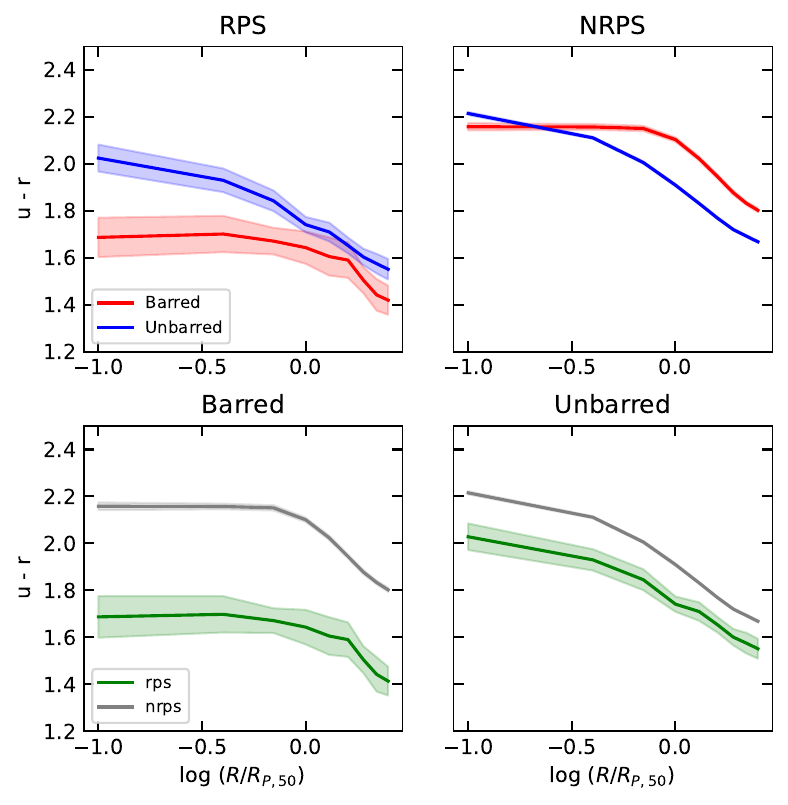}
    \caption{Unconstrained $u-r$ colour radial profiles as a function of the logarithmic normalized radius, log($R/R_{\mathrm{P,50}}$), where $R_{\mathrm{P,50}}$ is the Petrosian radius enclosing 50\% of the total flux. Red and blue symbols represent barred and unbarred galaxies, respectively. Green symbols represent RPS galaxies, while black symbols represent NRPS galaxies. \textit{Top left panel} includes all RPS galaxies and \textit{top right panel} all NRPS galaxies, both separated into barred and unbarred. In the \textit{lower-left panel}, we separate barred galaxies into RPS and NRPS, and in the \textit{lower-right panel} we repeat the same for unbarred galaxies. Each profile shows the median $u-r$ values in successive radial bins, with dispersion corresponding to bootstrap $1\sigma$ uncertainties.}
    \label{fig:rad_prof_all}
\end{figure}

Figure \ref{fig:rad_prof_all} displays the $u-r$ color radial profiles without considering differences in stellar mass and environmental conditions among the galaxies used to generate the radial profile for each subset. This figure serves as a baseline comparison to identify general trends before accounting for these factors. 

The top-left panel compares the  $u-r$ colour profiles of barred and unbarred galaxies affected by RPS. On global scales, both populations show negative colour gradients, with redder colours toward the centre and bluer colours at larger radii. The key difference appears in the central regions, where unbarred galaxies display a steeper central slope and become significantly redder toward the nucleus. In contrast, barred galaxies exhibit a flatter profile, showing comparatively bluer colours in their central regions. At larger radii, the two profiles gradually converge and reach similar colour values in the outer parts. This central difference in slope is visually apparent and will be examined in more detail in the following sections.

The top-right panel shows the $u-r$ colour profiles of barred and unbarred NRPS galaxies. Both profiles exhibit overall negative gradients, becoming redder toward the centre and bluer toward larger radii. The two populations overlap in the inner regions (log($R/R_{\mathrm{P,50}}$) $\leq -0.5$), but their profiles diverge at larger radii, where barred galaxies remain redder than unbarred ones. The shape of the profiles also differs since unbarred galaxies show a steady decline toward bluer colours with increasing radius, whereas barred galaxies present a central flattening followed by a milder decline. 

The bottom-left panel displays the $u-r$ colour profiles of barred galaxies, separated into RPS and NRPS subsets. Both profiles show overall negative gradients, and the RPS galaxies are systematically bluer across the entire radial range. The central regions of both profiles are relatively flat, with a mild decline beyond log($R/R_{\mathrm{P,50}}$) $\sim$ 0.0. The similarity in the shape of the profiles reinforces a trend already noted in the previous panels for barred galaxies, where central flattening combined with outer declines appears as a common feature. The main distinction here is the consistent offset in colour between the two populations.

The bottom-right panel displays the $u-r$ colour profiles of unbarred galaxies, separated into RPS and NRPS subsets. Both profiles show negative colour gradients and follow a broadly similar shape across the radial range. As in the previous panel, RPS galaxies are systematically bluer than their NRPS counterparts, although the colour offset between the two populations is less pronounced. The overall shapes remain comparable, with  steeper slopes than those observed in the barred galaxy samples.

In summary, the unconstrained $u-r$ colour profiles reveal systematic differences in both colour and shape between barred and unbarred galaxies, as well as between RPS and NRPS populations. Barred galaxies consistently show flatter profiles in the central regions, whereas unbarred galaxies display steeper gradients. Differences in overall colour are also present, although their interpretation requires caution at this stage. In the following sections, we refine this analysis by progressively controlling for stellar mass and environmental conditions in order to better isolate the effects of bars and RPS.

\subsubsection{Mass-matched Color Radial Profiles} \label{subsec: mass_crp}

\begin{figure}
	\includegraphics[width=\columnwidth]{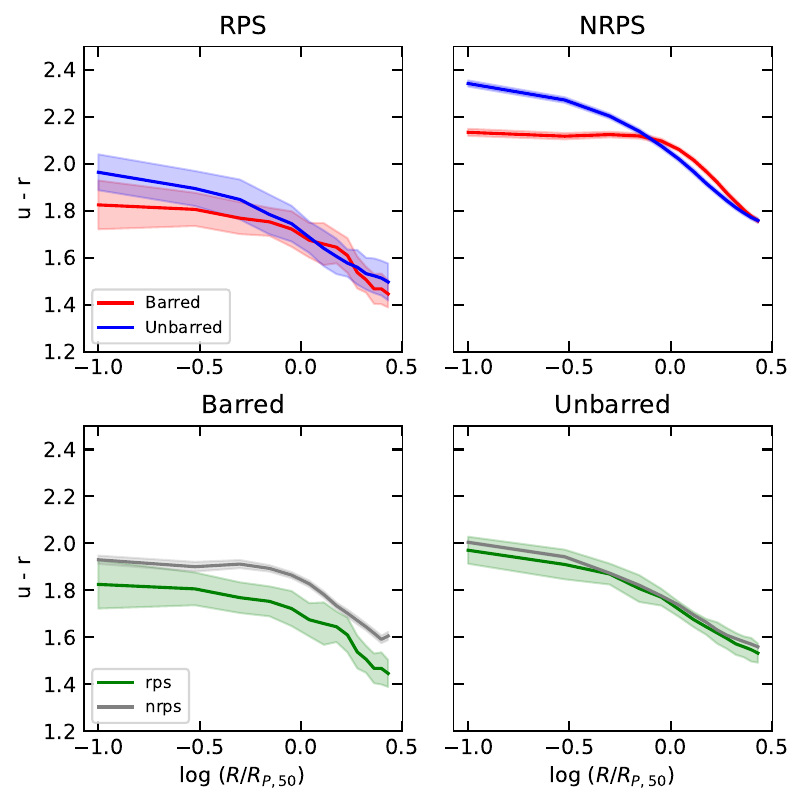}
    \caption{Radial $u-r$ colour profiles for the same galaxy subsets shown in Figure~\ref{fig:rad_prof_all}, but here the samples are controlled for stellar mass, as described in Section \ref{control by mass}.}
    \label{fig:rad_prof_mass}
\end{figure}

As a next step, we construct $u-r$ colour profiles from subsamples matched in stellar mass. In this step, the stellar mass distributions of the galaxies contributing to each radial profile are made comparable to those of their counterparts (see Section~\ref{control by mass}). This procedure reduces potential biases introduced by mass differences, allowing us to better isolate the effects of stellar bars and RPS. The results are presented in Figure~\ref{fig:rad_prof_mass}, which follows the same classification scheme as Figure~\ref{fig:rad_prof_all}, but now incorporates stellar mass matching.

Compared to the unconstrained case, mass matching reduces the absolute colour offsets between the different subsets, indicating that most of the differences previously observed were driven by underlying stellar mass variations. This behaviour is consistent with the well-known correlation between stellar mass, colour, and star formation activity, where more massive galaxies tend to host older stellar populations, exhibit redder colours, and form stars less efficiently than their lower-mass counterparts \citep[e.g.][]{LaraLopez+2013,Salim+2014,Schawinski+2014,Sanchez+2020}. In general, barred galaxies still exhibit flatter central colour profiles, while unbarred systems show steeper gradients. This suggests that bars help redistribute material within the disk and maintain central star formation, leading to smoother radial colour variations, a trend that persists even after controlling for stellar mass.

The top panels of Figure~\ref{fig:rad_prof_mass} show the $u-r$ colour profiles for barred and unbarred galaxies with matched stellar mass distributions, separated into RPS (left) and NRPS (right) populations. The most evident contrast occurs in the central regions, although subtle differences are also present in the outskirts. In both cases, the central region ($\log (R/R_{\mathrm{P,50}}) < 0.0$) reveals a consistent pattern in which barred galaxies display flatter profiles with comparatively bluer central colours, while unbarred galaxies decline more steeply toward redder values. These trends are consistent with previous studies suggesting that bars funnel gas toward the inner regions, thereby sustaining or enhancing central star-formation activity \citep[e.g.][]{Athanassoula+1992,Sheth+2005,Wang+2012,Lin+2017,Chown+2019,Lin+2020}.

At larger radii, the two profiles approach each other and follow similar trends. However, in the RPS sample, the outer regions are nearly indistinguishable between barred and unbarred galaxies, suggesting that the outskirts may be predominantly shaped by the stripping process rather than by bar-driven dynamics. In contrast, in the NRPS population, barred galaxies appear slightly redder in the outer regions, in line with the idea that bars tend to occur in more massive, intrinsically redder galaxies hosting older stellar populations \citep[e.g.][]{Masters+2012,Lee+2012,Kim+2017,Fraser-McKelvie+2020}.

When we compare only barred galaxies, split into RPS and NRPS populations in the bottom-left panel, RPS galaxies appear systematically bluer at all radii compared to their NRPS counterparts. In contrast, the bottom-right panel, which shows unbarred galaxies divided by RPS status, reveals no substantial difference in the radial colour profiles of the two groups. The overall shapes are similar in both cases, with the main distinction being the uniform offset toward bluer colours seen only in the barred sample.

This systematic offset in RPS galaxies (bottom-left panel) is consistent with previous observational results  which have reported enhanced star formation during the active phase of ram pressure stripping \citep[e.g.][]{Poggianti+2017, Vulcani+2018}. Similar trends have also been found in simulations, where the interaction between the galaxy and the intracluster medium compresses the gas and triggers star formation before quenching becomes dominant \citep[e.g.][]{Kronberger+2008,Steinhauser+2016}. However, the absence of a comparable offset in unbarred galaxies suggests that bars may modulate how RPS-induced star formation is distributed across the disk, enhancing its impact in the central regions.

Taken together, these results demonstrate that the observed differences in profile shape are not merely a by-product of mass variations but rather reflect the intrinsic effects of bars and RPS. In the next step of our analysis, we incorporate environmental conditions to further isolate the role of external processes in shaping the observed trends.

\subsubsection{Mass- and Environment-matched Colour Radial Profiles} \label{subsec: mass_env_crp}

\begin{figure}
	\includegraphics[width=\columnwidth]{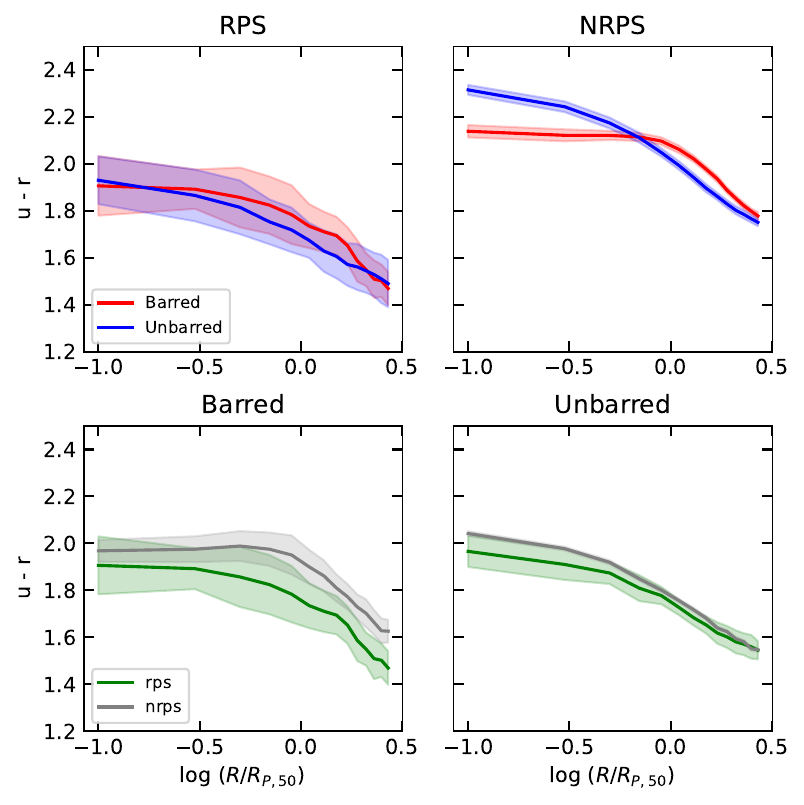}
    \caption{Radial $u-r$ colour profiles for barred and unbarred galaxies, separated into RPS and NRPS subsets, as in Figure~\ref{fig:rad_prof_all}. In this case, the samples are matched in both stellar mass and environment, as described in Section \ref{control by environment}.}
    \label{fig:rad_prof_mass_env}
\end{figure}

Figure~\ref{fig:rad_prof_mass_env} shows the radial $u-r$ colour profiles for galaxy samples matched simultaneously in stellar mass and environment, where the latter is defined by the position of each galaxy in phase space. By matching both stellar mass and environment, we ensure that differences in the radial colour profiles primarily trace the effects of bars, while minimising potential biases introduced by stellar mass or environmental variations.

The top-left panel shows that barred and unbarred galaxies in RPS pupulations have overall similar colour profiles across all radii. Nonetheless, barred systems retain a slightly flatter central slope, indicating that while the absolute colour differences largely vanish once stellar mass and environment are controlled, the structural imprint of the bar remains detectable in the profile shape. This indicates that the previously observed trend is at least partially influenced by external factors, such as the stage or duration of RPS experienced by each galaxy, rather than being driven solely by the intrinsic effect of bars.

The top-right panel shows that barred and unbarred NRPS galaxies display trends consistent with those found when controlling only for stellar mass (Figure~\ref{fig:rad_prof_mass}). As in the mass-matched case, barred galaxies exhibit flatter central profiles and are bluer in the inner regions compared to the steeper reddening of unbarred systems, consistent with previous findings of bar-driven central rejuvenation \citep[e.g.][]{Lin+2017, Lin+2020, Chown+2019}. In the outskirts, barred galaxies tend to be redder, although the contrast is weaker than in the centre. The fact that this trend persists even after simultaneously controlling for stellar mass and environment indicates that the NRPS comparison sample, although volume-limited and occupying similar phase-space positions to the RPS galaxies, is not strongly biased by the possible inclusion of systems with mild stripping signatures. In contrast, the top-left panel shows that for galaxies with clear optical signs of RPS, the central blue offset weakens significantly once environmental effects are accounted for, highlighting the dominant role of external processes in shaping their colour profiles.

The bottom-left panel shows that barred galaxies affected and unaffected by RPS still exhibit systematic differences, with RPS-barred systems remaining bluer than their NRPS counterparts across most radii. However, the contrast is weaker than in the previous analyses, particularly in the central regions where the confidence intervals overlap more substantially, underscoring the role of environmental conditions in modulating these trends. In the case of unbarred galaxies (bottom-right panel), a modest central difference becomes more apparent than in the mass-matched analysis, while the outer profiles remain largely indistinguishable. This emerging central offset may reflect variations in the stripping stage experienced by individual galaxies, despite having controlled for environmental effects, which could differentially affect their star formation histories. Nevertheless, when considering these limitations, the colour differences are more pronounced in barred than in unbarred systems, highlighting the stronger role of bars in shaping the impact of RPS.

These results indicate that, once both stellar mass and environment are controlled, the blue offset previously seen in barred RPS galaxies becomes less pronounced, suggesting that external processes play a significant role in shaping central colour variations during stripping. Even under these conditions, barred galaxies retain flatter central gradients than their unbarred counterparts, consistent with bar-driven inflows sustaining or enhancing central star formation. Overall, the colour structure of cluster galaxies appears to be primarily governed by stellar mass and environmental factors, with additional modulation by internal processes such as bar-driven inflows. The remaining variations likely depend on the specific stage of the stripping process.

\section{Summary and Conclusions} \label{sec: conclusions}

In this work, we investigated the combined effects of stellar bars and RPS on disk galaxies by building radial $u-r$ colour profiles from SDSS imaging. Galaxies were classified according to the presence of a stellar bar and signatures of RPS, and compared against a control NRPS population. To minimise biases, we applied a stepwise matching strategy in stellar mass and environment, allowing us to isolate the individual and combined imprints of bars and RPS on stellar population gradients.

As a first-order result, we find that all colour profiles in our analysis show negative $u-r$ gradients, with redder centres and bluer outskirts. This behaviour is consistent with the canonical inside–out formation of disk galaxies, where star formation propagates outward as galaxies evolve \citep[e.g.][]{White+1978,Wang+2011,Nelson+2016,Tacchella+2015}. Building on this framework, we find that bars and RPS introduce additional, superimposed imprints on the colour profiles. The most pronounced differences are seen when colour profiles are compared without any control for stellar mass or environment. RPS galaxies are systematically bluer than NRPS galaxies, and barred systems also appear bluer in their central regions than unbarred ones. These differences become less pronounced and, in some cases, even change their radial trend once stellar mass is controlled, indicating that part of the initial contrast is driven by the well-established dependence of colours and star formation rates on stellar mass, with more massive galaxies being redder and forming fewer stars per unit mass than their lower-mass counterparts \citep[e.g.][]{LaraLopez+2013,Salim+2014,Schawinski+2014,Sanchez+2020}.

After controlling for stellar mass, both RPS and NRPS populations show that barred galaxies are systematically bluer in their central regions than their unbarred counterparts, consistent with bars funneling gas inward and triggering enhanced nuclear star formation \citep[e.g.][]{Athanassoula+1992,Sheth+2005,Wang+2012,Lin+2017,Chown+2019,Lin+2020}. In RPS galaxies, the outer colour profiles of barred and unbarred systems are nearly indistinguishable, suggesting that the external regions are primarily shaped by the stripping process rather than by bar-driven dynamics. In NRPS systems, however, barred galaxies exhibit slightly redder outer regions, reflecting the tendency of bars to occur in more massive, intrinsically redder galaxies that host older stellar populations \citep[e.g.][]{Masters+2012,Lee+2012,Kim+2017,Fraser-McKelvie+2020}. Within the barred population, RPS galaxies are systematically bluer at all radii compared to their NRPS counterparts, while unbarred galaxies show no significant differences between the two groups. This indicates that bars play a key role in shaping the response to the cluster environment, as only barred systems translate the effects of ram pressure into a measurable rejuvenation of their stellar populations.

Once both stellar mass and environment are matched, colour offsets among the RPS galaxies become weaker but not entirely erased. Barred RPS systems still retain flatter central colour gradients than their unbarred counterparts, showing that the structural imprint of the bar persists even after controlling for internal and external factors. This indicates that the previously observed blue excess in barred RPS galaxies was partly driven by environmental differences, yet the intrinsic bar-induced rejuvenation remains detectable. In contrast, NRPS barred and unbarred galaxies display radial profiles with the same overall shape as in the mass-matched case, confirming that the NRPS set is not biased by galaxies with mild stripping signatures and can therefore serve as a reliable comparison baseline. When the samples are instead divided by bar presence, barred RPS galaxies remain globally bluer than their non-RPS counterparts, although the contrast weakens toward the centre, highlighting the role of environmental conditions in modulating but not erasing bar-driven activity. Unbarred galaxies, on the other hand, show only subtle central differences between RPS and NRPS systems.

Taken together, these results support a scenario in which stellar bars and ram pressure stripping can operate simultaneously within dense environments, consistent with our previous findings based on spatially and spectroscopically resolved data \citep{SanchezGarcia+2023}. Each mechanism can in principle induce gas inflow towards the center, bars by redistributing angular momentum and funneling gas toward the centre \citep[e.g.][]{Athanassoula+1992,Sakamoto+1999,Sheth+2005,Huang+2025,Yamamoto+2025}, and RPS through ISM–ICM mixing and pressure–gradient torques \citep[e.g.][]{Ramos-Martinez+2018,Akerman+2023,Kurinchi-Vendhan+2025}. The synergy between these two processes appears to favour temporary central rejuvenation in galaxies where gas compression by the intracluster medium can be efficiently channelled by the bar. Unbarred galaxies, by contrast, show no significant differences between RPS and NRPS populations when only stellar mass is controlled, but a mild central offset emerges once environment is also matched. This likely reflects a dependence on the stripping stage, since galaxies in our sample are not necessarily experiencing RPS at the same phase, and those observed at different stages of gas removal may display either enhanced or suppressed central activity depending on the remaining gas and duration of the interaction \citep[e.g.][]{Steinhauser+2016,George+2025}. Nevertheless, despite these possible differences in stripping stage, the rejuvenation signal becomes more evident when a bar is present to channel the gas compressed by RPS, producing the differences seen in the mass-matched case, which are largely diminished once environment is also controlled. This highlights the role of internal structures in modulating the observable outcome of environmental processes.

In line with this scenario we find that barred galaxies consistently display flatter central regions in their colour profiles compared to unbarred counterparts, regardless of whether stellar mass and environment are controlled. This persistent flattening across all configurations indicates that the bar-driven signature is a robust structural effect, even in dense environments where external processes such as ram pressure stripping also operate. Its apparent strength may, however, vary with the stage and intensity of stripping. Bars can funnel gas inward, sustaining central star formation and producing bluer cores \citep[e.g.][]{Athanassoula+1992,Sheth+2005,Lin+2017,Chown+2019,Lin+2020}, or alternatively redistribute material within the bar region, mixing stellar populations and inducing radial migration \citep[e.g.][]{Neumann+2020,Iles+2024,Bernaldez+2025}. In contrast, ram pressure alone may not strongly alter the radial distribution of stellar populations. In the absence of a bar, the interaction with the intracluster medium likely acts more uniformly across the disc, stripping or redistributing gas without efficiently funnelling it toward the centre, although this could depend on the stripping stage, which is not incorporated in this analysis.

Together with the evidence for flatter central profiles, these results underline the persistent role of bars in shaping the stellar populations of disk galaxies. We conclude that stellar bars and ram pressure stripping can act together within dense environments, producing more pronounced signatures of central rejuvenation than those observed in systems experiencing ram pressure alone. In this framework, bars play a key role in channelling or redistributing gas compressed by the intracluster medium, thereby amplifying the observable impact of environmental processes. While other mechanisms operating in dense environments may produce similar effects, our results indicate that the interplay between bars and ram pressure represents an additional and potentially important pathway for galaxy evolution in clusters. Future work is required to test this scenario more broadly by constructing larger RPS samples that allow the study of different stages of ram pressure stripping, and by exploring complementary tracers of star formation and gas content. An important avenue will be to investigate whether the combined effect of bars and RPS is connected to enhanced nuclear activity, and to assess how environmental processes may influence bar properties themselves, such as bar length and strength.

\section*{Acknowledgements} \label{sec: acknowlwdgements}

The authors thank the anonymous referee for useful comments that helped to improve the quality of the paper and clarify our results. KMD thanks the support of the Serrapilheira Institute (grant Serra-1709-17357) as well as that of the Brazilian National Research Council (CNPq grant 308584/2022-8) and of the Rio de Janeiro Research Foundation (FAPERJ grant E-32/200.952/2022), Brasil. OSG and BCS acknowledge the financial support provided by PAPIIT project IN111825 from DGAPA-UNAM.\\

\textit{Software}: {\sc Astropy} \citep{Astropy+2013, Astropy+2018, Astropy+2022}, {\sc MatchIt} \citep{Ho+2011}, {\sc matplotlib} \citep{Hunter+2007}, {\sc numpy} \citep{Harris+2020}, {\sc pandas} \citep{Reback+2020}, {\sc photutils} \citep{Bradley+2025}, {\sc R} \citep{R+2015}, {\sc scipy} \citep{Virtanen+2020}, {\sc seaborn} \citep{Waskom+2021}, and {\sc topcat} \citep{Taylor+2005}.

\section*{Data Availability}
The data underlying this article were collected public data. Visual bar classifications for the RPS sample will be shared on reasonable request to the corresponding author.






\bsp	
\label{lastpage}
\end{document}